\documentstyle[aps,prl]{revtex}
\begin{document}
\narrowtext
\twocolumn
\pagestyle{empty}

\noindent{\bf Comment on ``Bicritical and Tetracritical
\newline Phenomena and Scaling Properties of the \newline SO(5) Theory''}

\bigskip

Recently, Hu~\cite{hu} used Monte Carlo simulations on an SO(5)
rotator model, and concluded that the multicritical point which
characterizes the simultaneous ordering of the SO(3)
``antiferromagnetic" (AF) 3-component and of the U(1)
``superconducting" (SC) 2-component order parameters,
$\overrightarrow S$ and $\overrightarrow \Delta$, has the critical
behavior of the {\it isotropic} 5-component rotator model. This
contradicts the order-$\epsilon$ renormalization group (RG) in
$d=4-\epsilon$ dimensions, which states that the isotropic SO(5)
fixed point is unstable, and identifies this multicritical point
with the anisotropic biconical fixed point~\cite{BL,kosterlitz}.
Measurements of isotropic 5-component critical exponents at this
multicritical point were proposed as ``measuring the number 5,
confirming the SO(5) theory for high-$T_c$
superconductivity~\cite{zhang}.

Without addressing the relevance of this discussion to high-$T_c$
superconductivity (where one should also include fluctuations in
the electromagnetic gauge field~\cite{ma}), I show here that in
fact, at $d=3$ the multicritical point {\it must} be
tetracritical, being characterized by the {\it decoupled} fixed
point, at which the two order parameters are asymptotically
independent. $\overrightarrow S$ and $\overrightarrow \Delta$
exhibit the Heisenberg ($n=3$) and XY ($n=2$) critical exponents
even when they order simultaneously, and the two critical lines
cross each other at finite angles. Accurate experiments in the
asymptotic regime thus carry no information on the SO(5) theory.

The stability of the decoupled fixed point follows from an exact
argument, which was already presented in 1976~\cite{rg,AF}: at
this point, the coupling term $|\overrightarrow
S|^2|\overrightarrow \Delta|^2$ scales like the product of two
energy-like operators, having the dimensions
$(1-\alpha_{n})/\nu_n$, where $\alpha_n$ and $\nu_n$ are the
specific heat and correlation length exponents. Thus, the combined
operator has the dimension $d-\lambda$, where
\begin{equation}
\lambda=\frac{1}{2}\big(\frac{\alpha_2}{\nu_2}+\frac{\alpha_3}{\nu_3}\big)
\end{equation}
is the scaling exponent which determines the RG flow of the
coefficient of this term near the decoupled fixed point. The known
negative values of $\alpha_2$ and $\alpha_3$ at $d=3$~\cite{legui}
then yield $\lambda\cong-0.087<0$, and the decoupled fixed point
is stable, in contrast to the order-$\epsilon$ extrapolation to
$\epsilon=1$~\cite{BL,kosterlitz}.

Ref. \onlinecite{hu} used a discrete spin model, with the
constraint $|\overrightarrow{S}|^2
+|\overrightarrow{\Delta}|^2=1$. This is believed to be in the
same universality class as a Ginzburg-Landau-Wilson (GLW) theory,
with the quartic term
$u(|\overrightarrow{S}|^2+|\overrightarrow{\Delta}|^2)^2$ (where
initially $u\longrightarrow\infty$)~\cite{WK}. Ref.
\onlinecite{hu} then added a coupling
$w|\overrightarrow{S}|^2|\overrightarrow{\Delta}|^2$. Quantum
fluctuations~\cite{hanke} and RG iterations~\cite{rg} then also
generate a term
$v(|\overrightarrow{S}|^4-|\overrightarrow{\Delta}|^4)$.
Experience~\cite{kosterlitz,rg} yields six fixed points in the
$u-v-w$ parameter space, of which {\it only one} is stable. For a
continuous transition, the above argument implies an RG flow away
from the vicinity of the isotropic SO(5) unstable fixed point, at
$v=w=0$, to the decoupled one, where $2u+w=0$. This flow may be
slow, since the related exponents are small. Therefore, one might
need to go very close to the tetracritical point in order to
observe the correct critical behavior. The simulations of Ref.
\onlinecite{hu}, which begin close to the isotropic fixed point
($u\gg v,w$) and use relatively small samples, probably stay in
the transient regime which exhibits the isotropic exponents. To
observe the true asymptotic decoupled behavior, one should start
with a more general model, allowing different interactions for
$\overrightarrow{S}$ and for $\overrightarrow{\Delta}$ and
relaxing the strong constraint $|\overrightarrow{S}|^2
+|\overrightarrow{\Delta}|^2=1$.

Finally, note the different scaling behavior near the decoupled
tetracritical point: the free energy breaks into a sum of the two
free energies, and the asymptotic crossover exponent is
$\varphi=1$. However, there are also slowly decaying corrections,
proportinal to $w\xi^{\lambda}$, where $\xi$ is some average
correlation length. All the above statements assume that the
initial Hamiltonian is the the region of attraction of the
decoupled fixed point. Alternatively, one might expect a first
order transition.

 This work was supported by the German-Israeli Foundation (GIF).

\bigskip
\noindent Amnon~Aharony \newline {\small School of Physics and
Astronomy, Tel Aviv University, Tel Aviv 69978, Israel.}

\bigskip
\noindent Received xxx \newline \noindent PACS numbers {74.20.-z,
05.70.Jk, 74.25.Dw }

\end{document}